\documentclass[a4paper,12pt]{article}
\usepackage{amsfonts}

\newcommand{\Ga}{\alpha}

\newcommand{\Gd}{\delta}
\newcommand{\Ge}{\epsilon}
\newcommand{\Gg}{\gamma}
\newcommand{\GG}{\Gamma}

\newcommand{\Gl}{\lambda}
\newcommand{\GL}{\Lambda}

\newcommand{\Gth}{\theta}
\newcommand{\GTh}{\Theta}

\newcommand{\cA}{{\scriptscriptstyle\cal A}}
\newcommand{\cB}{{\scriptscriptstyle\cal B}}
\newcommand{\cC}{{\scriptscriptstyle\cal C}}
\newcommand{\cD}{{\scriptscriptstyle\cal D}}
\newcommand{\CD}{{\cal D}}

\newcommand{\cL}{{\cal L}}

\newcommand{\cO}{{\cal O}}
\newcommand{\cP}{{\cal P}}
\newcommand{\cQ}{{\cal Q}}

\newcommand{\cV}{{\cal V}}

\newcommand{\dA}{{\dot{A}}}
\newcommand{\dB}{{\dot{B}}}
\newcommand{\dC}{{\dot{C}}}

\newcommand{\ft}[2]{{\textstyle {\frac{#1}{#2}} }}

\newcommand{\dd}{\partial}

\newcommand{\ra}{\rightarrow}
\newcommand{\I}{{\rm i}}

\newcommand{\be}{\begin{equation}}
\newcommand{\ee}{\end{equation}}
\newcommand{\ben}{\begin{displaymath}}
\newcommand{\een}{\end{displaymath}}
\newcommand{\ba}{\begin{eqnarray}}
\newcommand{\ea}{\end{eqnarray}}
\newcommand{\nn}{\nonumber}
\newcommand{\non}{\nonumber\\}

\newcommand{\la}{\label}
\newcommand{\ci}{\cite}

\newcommand{\Ref}[1]{(\ref{#1})}

\def\moth{\mathsurround=0pt}
\newdimen\zo \zo=0pt

\def\tick{\leaders\hrule height 0.5ex depth 0pt \hskip 0.5pt}
\def\upboxfill{$\moth \setbox\zo\hbox{\tick}%
  \hskip 2pt\hbox to 0pt{$\tick$\hss}\hrulefill \hbox to 6pt{$\tick$\hss}$}
\def\underbox#1{\offinterlineskip{\mathord{\mathop{\vtop{\moth\ialign{##\crcr
      $\hfil\displaystyle{#1}\hfil$\crcr\noalign{}
      {\upboxfill}\crcr\noalign{}}}}\limits}}}
\def\dtick{\leaders\hrule height .34pt depth .5ex \hskip 0.5pt}
\def\downboxfill{$\moth \setbox\zo\hbox{\dtick}%
  \hskip 2pt\hbox to 0pt{$\dtick$\hss}\hrulefill \hbox to 6pt{$\dtick$\hss}$}

\def\undersym#1{\underbox{{}#1}}

\newcommand{\vl}{{\vphantom{[}}}
\newcommand{\VV}[2]{{\cV^{\,#1}{}\!^\vl_{#2}}}

\begin{document}

\thispagestyle{empty}

\begin{flushright}
AEI-2000-065\\
LPTENS-00/36\\
hep-th/0010076
\end{flushright}
\renewcommand{\thefootnote}{\fnsymbol{footnote}}

\vspace*{0.1cm}
%\noindent
%\rule{\linewidth}{0.6mm}
%\vspace*{0.6cm}
\begin{center}
{\bf\Large Maximal gauged supergravity}
\medskip

{\bf\Large in three dimensions}

\bigskip\bigskip\medskip

{\bf H.~Nicolai\medskip\\ }
{\em Max-Planck-Institut f{\"u}r Gravitationsphysik,\\
  Albert-Einstein-Institut,\\
  M\"uhlenberg 1, D-14476 Potsdam, Germany}

\smallskip {\small nicolai@aei-potsdam.mpg.de} \bigskip\smallskip

{\bf H.~Samtleben\medskip\\ }
{\em Laboratoire de Physique Th{\'e}orique\\
  de l'Ecole Normale Sup{\'e}rieure,
\footnote{UMR 8548: Unit{\'e} Mixte du CNRS et de l'ENS. }\\
  24 Rue Lhomond, F-75231 Paris Cedex 05, France}~
\\

\smallskip {\small henning@lpt.ens.fr\medskip} 
\end{center}
\renewcommand{\thefootnote}{\arabic{footnote}}
\setcounter{footnote}{0}
\bigskip
\medskip

\begin{abstract}
We construct maximally supersymmetric gauged $N\!=\!16$ supergravity
in three dimensions, thereby obtaining an entirely new class of 
AdS supergravities. These models are not derivable from any known 
higher-dimensional theory, indicating the existence of a new type of 
supergravity beyond $D\!=\!11$. They are expected to be of special 
importance also for the conjectured AdS/CFT correspondence. One of 
their noteworthy features is a nonabelian generalization of the 
duality between scalar and vector fields in three dimensions.
Among the possible gauge groups, $SO(8)\!\times\!SO(8)$ is distinguished 
as the maximal compact gauge group, but there are also more exotic 
possibilities such as $F_{4(-20)}\!\times\!G_2$.
\end{abstract}

\renewcommand{\thefootnote}{\arabic{footnote}}
\vfill
\leftline{{ October 2000}}

\setcounter{footnote}{0}
\newpage

Gauged supergravities have attracted strong interest again recently in
the context of the conjectured duality between AdS supergravities and
superconformal quantum field theories on the AdS boundary
\cite{AGMOO00}.  Since the construction of maximal $N=8$ gauged
supergravity in four dimensions \ci{deWNic82} various maximal gauged
supergravities in dimensions $D\!>\!4$ have been shown to exist
\ci{SalSez89}. Much less is known, however, about gauged
supergravities in dimensions $D<4$ --- despite the fact that the
theories admitting AdS$_3$ ground states are expected to be of
particular interest for the AdS/CFT duality due to the rich and rather
well understood structure of two-dimensional superconformal field
theories. A better understanding of the supergravity side is of vital
importance for any further test of this duality.

In this letter, we construct maximally supersymmetric gauged
$N\!=\!16$ supergravity in three dimensions. This is the most complex 
and most symmetric of all theories of this type known so far (and 
may remain so in view of the appearance of the maximal exceptional 
Lie group $E_{8(8)}$ in it). It gives rise to an entirely new class 
of three-dimensional models, none of which is contained in or derivable 
by any known mechanism from the known supergravity or superstring theories 
in higher dimensions. This is all the more remarkable in view of the fact 
that these models constitute continuous deformations of a theory (ungauged
$N\!=\!16$ supergravity \cite{Juli83,MarSch83}) that {\it is} obtained by a
torus compactification of $D\!=\!11$ supergravity \cite{CrJuSc78}.
Their existence not only hints at a new type of supergravity beyond 
$D\!=\!11$ supergravity, but also indicates that the framework of 
theories considered so far in attempts to unify string theories at 
the non-perturbative level may be inadequate.
Likewise, the boundary theory associated with our
$SO(8)\!\times\!SO(8)$ gauged supergravity is expected to enjoy an
exceptional status among the superconformal field theories. The
AdS/CFT correspondence indeed suggests a duality with maximally
supersymmetric Yang-Mills theory on the two-dimensional AdS boundary
--- but this is just matrix string theory, which itself has
been proposed as a candidate for M theory \cite{DeVeVe97}!

Topological gauged supergravities in three dimensions were constructed
in \ci{AchTow86}; these theories are supersymmetric extensions of
Chern-Simons (CS) theories with $(p,q)$ supersymmetry and gauge groups
$O(p)\!\times\!O(q)$, but have no propagating matter degrees of
freedom. Matter coupled gauged supergravities can, of course, be
obtained by direct dimensional reduction of gauged supergravities in
$D\!\geq\!4$ to three dimensions and below \cite{LuPoTo97}; see also
\cite{CvLuPo00} and \cite{DKSS00} for different constructions. However, 
no such theories preserving the maximal number of supersymmetries 
in three dimensions have been known until now.

The scarcity of results is in part explained by the different status
of vector fields in three dimensions in comparison with their higher
dimensional relatives, namely their on-shell equivalence with scalar
fields via the duality relation $\Ge_{\mu\nu\rho}\, \dd^\rho \varphi^m
= \dd_{[\mu} {B_{\nu]}}^m $. This relation plays a special role in the
derivation of maximal $N\!=\!16$ supergravity in three dimensions
\ci{Juli83,MarSch83}: in order to expose its rigid $E_{8(8)}$ 
symmetry, all vector fields obtained by dimensional reduction of 
$D\!=\!11$ supergravity \ci{CrJuSc78} on an 8-torus must be dualized
into scalar fields. Thus an immediate (but, as we will see, only 
apparent) obstacle towards the gauging of the maximally extended 
three-dimensional supergravity is the {\em a priori} absence
of vector fields that could be used for such a gauging. Let us
therefore emphasize the main features by which our construction
differs from previous ones:

\begin{itemize}

\item The theory makes simultaneous use of both scalar fields and
their dual vectors, which appear via a CS-term in order of the gauge 
coupling constant $g$ and therefore do not carry additional physical 
degrees of freedom. Their equations of motion yield a non-abelian
generalization of the standard duality relation between scalar and vector 
fields, such that a direct elimination of the latter is no longer possible,
and would lead to a non-local theory in terms of the 128 physical 
scalar fields of $N\!=\!16$ supergravity. 

\item Several consistent choices are possible for the gauge group
$G_0\subset E_{8(8)}$, although $G_0= SO(8)\!\times\!SO(8)$ is
distinguished as the maximal compact subgroup of $E_{8(8)}$ that can
be made local. The non-compact gauge groups allowed by
%The associated gauged supergravity contains the $(8,8)$
%CS-theory of \cite{AchTow86}. 
our construction do not arise as analytic continuations of a compact
subgroup unlike the known non-compact gaugings in higher dimensions
\cite{Hull84}.

\item Conventional Kaluza Klein compactification of $D\!=\!11$
(or $D\!=\!10$) supergravity cannot give rise to these gauged 
theories because ($i$) it would yield Yang-Mills type theories 
rather than non-abelian CS theories, and ($ii$) there are no 
8-manifolds (or 7-manifolds) whose isometry groups would coincide 
with any of the possible gauge groups $G_0$.

\item Although the rigid $E_{8(8)}$ symmetry is broken, such that the
gauged theory is only invariant under local $G_0 \times SO(16)$, the
construction makes essential use of properties of the exceptional Lie
group $E_{8(8)}$. In particular, the check of local supersymmetry
requires ``$T$-identities'' based on $E_{8(8)}$ analogous to the ones
derived in \cite{deWNic82}.

\end{itemize}

The physical fields of ungauged maximal $N\!=\!16$ supergravity in
three dimensions constitute an irreducible supermultiplet with 128
bosons and 128 fermions transforming as inequivalent fundamental
spinors of $SO(16)$ (see \ci{MarSch83} whose conventions and notation
we follow). In addition, the theory contains the dreibein
${e_\mu}^\alpha$ and 16 gravitino fields $\psi_\mu^I$. The scalar
fields are described by an element $\cV$ of the non-compact coset
space $E_{8(8)}/SO(16)$ in the fundamental 248-dimensional
representation \cite{Juli83}. The equivalence of the fundamental and
the adjoint representations of $E_{8(8)}$ is expressed by the relation
\be\la{adjoint}
\cV^{-1} t^\cA\, \cV = {\cV^\cA}_\cB \,t^\cB \;.
\ee
Indices ${\scriptstyle\cal{A}}, {\scriptstyle\cal B}, \dots$ here
label $E_{8(8)}$ generators $t^\cA$ (see \ci{KoNiSa99a,KoNiSa99b} for 
our $E_{8(8)}$ conventions and some useful formulas). We split the
generators $\{t^\cA\}$ into 120 compact ones $\{X^{IJ}\}$ and 128
noncompact ones $\{Y^A \}$. 

$N\!=\!16$ supergravity is invariant under a hidden rigid $E_{8(8)}$
symmetry as well as a local $SO(16)$ symmetry. The supersymmetry 
variations are \cite{MarSch83}
\be\la{susyf}
\begin{array}{rclcrcl}
\delta {e_\mu}^\Ga &=& \I \bar \Ge^I \gamma^\Ga \psi_\mu^I \;,
&\;\;&
\cV^{-1} \delta \cV &=& \Gamma^I_{A\dot A} \bar \chi^{\dot A} \Ge^I Y^A
\;,\\[1ex]
\delta\,\psi^I_\mu &=& D_{\mu}\epsilon^I \;,
&&
\Gd\,\chi^\dA &=&
\ft{\I}2\,\Gg^\mu\Ge^I\,\GG^I_{A\dA}\,P_\mu^A \;.
\end{array}
\ee
The scalar fields couple to the fermions via the currents
\be\la{VdV}
\cV^{-1} \dd_\mu \cV = \ft12 Q_\mu^{IJ} X^{IJ}\!+\!P_\mu^A Y^A \;.
\ee
where the composite $SO(16)$ connection $Q_\mu^{IJ}$ enters the
covariant derivative $D_\mu$ in \Ref{susyf} and \Ref{Lagrangian} below.

A central role in our construction is played by the supercovariant 
duality relation
\ba\la{duality}
&&\Ge^{\mu\nu\rho}\,\Big(
B_{\nu\rho}{}^m + 2\VV{m}{IJ}\,\overline{\psi}{}^I_\nu\psi^J_\rho
-2\I\,\GG^I_{A\dA}\VV{m}{A}\,\overline{\psi}{}^I_\nu\Gg_\rho\chi^\dA
\Big) \la{dual}\\
&&=2\VV{m}{A}\Big(P^{\,\mu A}
-\GG^I_{A\dA}\overline{\psi}{}^{\,\mu\,I}\chi^\dA \Big)
-\ft{\I}2\,\VV{m}{IJ}\,\GG^{IJ}_{\dA\dB}\,
\overline{\chi}^\dA\Gg^\mu\chi^\dB \;,
\nn
\ea
whose compatibility condition gives rise to the equations of motion
for the scalar fields. At the same time \Ref{duality} serves to define
the dual vector fields ${B_\mu}^m$ with (abelian) field strengths
$B_{\mu\nu}{}^m$. Although we are still concerned with abelian vector
fields at this point, we introduce special labels $m,n,\dots$ for the
generators of the gauge group $G_0\subset E_{8(8)}$ already here. The
gauge group is properly described by means of its embedding tensor
$\Theta_{\cA\cB}$, such that
\be\la{m}
{B_\mu}^m t_m \equiv {B_\mu}^\cA \Theta_{\cA\cB} \, t^\cB \;.
\ee
We emphasize that the number of vector fields is not yet specified
since it will depend on the choice of the gauge group $G_0$. 
%Trading part of the scalar fields for vector fields by means of 
%\Ref{duality} gives rise to various equivalent formulations of the 
%ungauged three-dimensional theory \cite{CJLP98}. 
Although the duality \Ref{dual} holds only on-shell, we shall see 
that the gauged theory provides a natural off-shell framework which 
accommodates both the scalars and their dual vectors.

Under local supersymmetry the vectors transform as~\cite{KoNiSa99b}
\be\la{susyv}
\Gd B_\mu{}^m ~=~ -2 \VV{m}{IJ}\,\overline{\Ge}^I\psi^I_\mu
+\I \GG^I_{A\dA}\,\VV{m}{A}\,\overline{\Ge}^I\Gg_\mu\chi^\dA \;.
\ee

To gauge the theory, we now assume the existence of a suitable
subgroup $G_0\subset E_{8(8)}$ with symmetric embedding tensor
$\GTh_{\cA\cB}$. The possible choices for $G_0$ will be determined
below by demanding consistency and full supersymmetry of the gauged
theory. The first step is the covariantization of derivatives in
\Ref{VdV} according to
\ba\la{gauging}
\cV^{-1}\,\CD_{\!\mu} \cV &\equiv&
\cV^{-1}\dd_\mu \cV 
+ g\,B_\mu{}^m \, \cV^{-1} t_m \, \cV \non[1ex] 
&=& \cP_\mu^AY^A +\ft12 \cQ_\mu^{IJ}X^{IJ} \;,
\ea
with gauge coupling constant $g$, where we now use projected indices
$m,n,\dots$ for the gauge group. The non-abelian field strength reads
\ba\la{FS}
B_{\mu\nu}{}^m &=& \dd_\mu\,B_{\nu}{}^m - \dd_\nu\,B_{\mu}{}^m
+ g\,f^{m}{}_{np}\,B_{\mu}{}^n B_{\nu}{}^p \;.
\ea
The modified currents defined in (\ref{gauging}) lead to a first
$g$-dependent modification of the ungauged Lagrangian \ci{MarSch83}
\ba\la{Lagrangian}
\cL^{(0)} &=& 
 -\ft14 e R
 + \ft14 e \cP^{\mu A} \cP^A_\mu
+\ft12\, \Ge^{\Gl\mu\nu} \overline{\psi}{}^I_\Gl D_\mu \psi_\nu^I 
\non[1ex]
&&{}
-\ft{\I}{2}e \overline{\chi}^{\dA} 
\Gg^\mu D_\mu \chi^{\dA} 
-\ft12e\,  \overline{\chi}^{\dA} 
\Gg^\mu \Gg^\nu \psi^I_\mu \,\GG^I_{A\dA} \cP^A_\nu \;.
\ea
Next we observe that the non-abelian extension of the duality 
relation \Ref{dual} is obtained from this Lagrangian if we add 
the following CS-term for the vector fields $B_\mu{}^m$:
\ba
\cL^{(1)}_g &=& 
-\ft14\,g\,\Ge^{\mu\nu\rho}\,B_\mu{}^m
\Big(\dd_\nu B_\rho\,{}_m
+\ft13\,g f_{mnp}\,B_\nu{}^n B_\rho{}^p \Big) 
\la{L1}
\ea
This is very different from the situation in higher dimensions where
the vector fields appear already in the ungauged theory via an abelian
kinetic term. By contrast, we here find that the CS-term \Ref{L1}
induces a {\em nonabelian} version of the duality relation \Ref{dual}
between scalar and vector fields; the latter is simply obtained by
replacing the abelian field strengths by \Ref{FS}, and the current
$P_\mu^A$ by the one defined in \Ref{gauging}, with no further
modifications. This shows how the gauged theory brings in the dual
vector fields, but without introducing new propagating degrees of
freedom. However, because of the explicit appearance of the gauge
fields on the r.h.s.\ of the non-abelian duality relation it is no
longer possible to trade the vector fields for scalar fields and
thereby eliminate them, unlike in \cite{CJLP98}.

The next steps in the construction are standard \ci{deWNic82}: the 
$\cO (g)$ terms in the currents necessitate extra bilinear fermionic 
terms in the Lagrangian
\ba
\cL^{(2)}_g &=& 
\ft12ge\,
A_{1}^{IJ}\;\overline{\psi}{}^I_{\mu}\,\Gg^{\mu\nu}\,\psi^{J}_{\nu} + 
\non
&&{}
+{\I}ge\,A_{2}^{I\dA}\;
\overline{\chi}{}^\dA\,\Gg^\mu\,\psi^I_{\mu}
+ \ft12ge\,
A_{3}^{\dA\dB}\;\overline{\chi}{}^\dA\,\chi^{\dB}\;,
\la{L2}
\ea
as well as the following modifications of the fermionic variations:
\be
\Gd_g \psi^I_\mu = \I g\,{A}_{1}^{IJ}\,\Gg_{\mu} \Ge^J \;, \qquad
\Gd_g \chi^\dA =  g\,A_{2}^{I\dA}\,\Ge^I \;.  \la{ferm}
\ee
Here the $SO(16)$ tensors $A_{1,2,3}$ depend on the scalar fields
$\cV$ in a way that remains to be specified, and introduce Yukawa-type 
couplings between the scalars and the fermions beyond the derivative
couplings generated by \Ref{VdV}. At $\cO (g^2)$, supersymmetry demands 
that we add the scalar potential
\ba\la{potential}
\cL^{(3)}_g &=& 
\ft18 \,g^2\,e\,\Big(
A_{1}^{IJ}A_{1}^{IJ}-\ft12\,A_{2}^{I\dA}A_{2}^{I\dA} \Big) \;.
\ea

The check of local supersymmetry requires several identities
that these tensors must obey. We here list only the most important ones.
They are
%Implicity, the tensors $A_{1,2,3}$ are defined by the relations
\ba
\VV{m}{A}\,\GG^{(I}_{A\dA}\,A_2^{J)\dA} &=&
\VV{m}{IK}\,A_1^{JK} + \VV{m}{JK}\,A_1^{IK}\;, 
\non
\GG^{[I}_{A\dA}\,A_2^{J]\dA} &=& \VV{\cC}{IJ}\GTh_{\cC\cD}\VV{\cD}{A} \;,
\non
\GG^I_{A\dA} \,A_3^{\dA\dB} \,\GG^I_{\dB B} &=&
16\, \VV{\cC}{A}\GTh_{\cC\cD}\VV{\cD}{B} \;,
\la{id1}
\ea
with $\GG^I_{A\dA} A_2^{I\dA} = 0$. In addition we have the 
differential relations
\ba
\CD_\mu A_{1}^{IJ} &=& \cP_\mu{}^A\,\GG^{(I}_{A\dA}\,A_2^{J)\dA} \;, 
\non
\CD_\mu A_{2}^{I\dA} &=& 
\ft12\,\cP_\mu{}^A\,
\Big(\GG^I_{A\dB}\,A_3^{\dA\dB} + \GG^J_{A\dA}\,A_1^{IJ} \Big) 
\non
&&{}-\ft12\,\cP_\mu{}^A\,
\GG^I_{B\dA}\,\VV{\cA}{A}\GTh_{\cA\cB}\VV{\cB}{B} \;,
\la{id2}
\ea
and the quadratic identities
\ba
&& A_1^{IK}A_1^{KJ} - \ft12\,A_2^{I\dA}A_2^{J\dA}
\non
&&\qquad\qquad = \ft1{16}\,\Gd^{IJ}\,
\Big(
A_1^{KL}A_1^{KL} - \ft12\,A_2^{K\dA}A_2^{K\dA}
\Big) \;,
\non[2ex]
&&3\,A_1^{IJ}A_2^{J\dA} - A_2^{I\dB}A_3^{\dA\dB}
\non
&&\qquad\qquad = \ft1{16}\,\GG^I_{A\dA}\,\GG^J_{A\dB}\,
\Big(3\,A_1^{JK}A_2^{K\dB} - 
A_2^{J\dC}A_3^{\dB\dC}\Big) \;.
\la{id3}
\ea
Remarkably, the $SO(16)$ representations of the tensors $A_{1,2,3}$
combine into a representation of $E_{8(8)}$ according to ${\bf
135}\oplus {\bf 1820} \oplus {\bf 1920} = {\bf 3875}$, with the
possibility of an extra singlet in the trace parts of $A_1^{IJ}$ and
$A_3^{\dA \dB}$. Apart from the extra singlet, this is analogous to
the situation in $D\!=\!4$ and $D\!=5$, where the Yukawa couplings
are given by tensors transforming in the ${\bf 912}$ of $E_{7(7)}$
\cite{deWNic84} and in the ${\bf 351}$ of $E_{6(6)}$ \cite{GuRoWa86},
respectively.

By a lengthy calculation we can now establish that the full Lagrangian
\be\la{L}
\cL = \cL^{(0)}+\cL^{(1)}_g+\cL^{(2)}_g+\cL^{(3)}_g \;,
\ee
is indeed supersymmetric provided the tensors $A_{1,2,3}$ solve
equations \Ref{id1}--\Ref{id3}. In the limit $g\!\ra\!0$ one
re-obtains the ungauged theory of ref.\cite{MarSch83}. Furthermore, 
the $N\!=\!16$ superalgebra closes in the usual way, except for an 
extra non-abelian $G_0$--gauge transformation with parameter
\be
\GL^m = 
2\,\VV{m}{IJ}\,\overline{\Ge}{}^I_1\Ge^J_2 
+\I B_\mu{}^m\,\overline{\Ge}{}^I_1\Gg^\mu \Ge^I_2 \;.
\ee

The main task is now to ascertain that the relations
\Ref{id1}--\Ref{id3} do admit nontrivial solutions, and to determine
the possible gauge groups $G_0$. Any subgroup $G_0\subset E_{8(8)}$ is
characterized by its embedding tensor $\GTh_{\cA\cB}$ (cf.\
\Ref{m}) which decomposes as
\be
\GTh_{\cA\cB} = \Gth\,\eta_{\cA\cB}
+ \GTh_{\cA\cB}^{\bf 3875} + \GTh_{\cA\cB}^{\bf 27000} \;,
\ee
with the (indefinite) $E_{8(8)}$ Cartan-Killing metric $\eta_{\cA\cB}$, 
in accordance with the symmetric $E_{8(8)}$ tensor product
\be
({\bf 248}\times{\bf 248})_{\rm sym} = 
{\bf 1}+{\bf 3875}+{\bf 27000} \;.
\ee
Guided by \cite{deWNic82} we define the $T$-tensor by
\be\la{T}
T_{\cA|\cB}\equiv \VV{\cC}{\cA}\VV{\cD}{\cB}\,\GTh_{\cC\cD} \;,
\ee
and note that equations \Ref{id1} define the $SO(16)$ tensors
$A_{1,2,3}$ as functions of $T_{\cA|\cB}$ (unlike the cubic
expressions in \cite{deWNic82} and \cite{GuRoWa86}, the $T$-tensor 
\Ref{T} is quadratic in $\cV$ due to \Ref{adjoint}). 
As the tensors $A_{1,2,3}$ combine 
into the ${\bf 3875}( + {\bf 1})$ representation(s) of $E_{8(8)}$, 
the equations \Ref{id1}--\Ref{id3} can only be consistent if the 
$\bf 27000$ is absent in \Ref{T}, which in turn requires $\GTh_{\cA\cB}$ 
to satisfy the condition
\be\la{crit}
\GTh_{\cA\cB}^{\bf 27000}= 0\quad\Longleftrightarrow\quad
\GTh_{\cA\cB} = \Gth\,\eta_{\cA\cB}
+ \GTh_{\cA\cB}^{\bf 3875} \;.
\ee
If \Ref{crit} is satisfied, the tensors $A_{1,2,3}$ which solve
equations \Ref{id1}--\Ref{id3} are given by
\ba
A_{1}^{IJ} &=& 
-\Gth\,\Gd_{IJ}
-\ft1{7}\,\VV{\cC}{IK}\VV{\cD}{KJ}\,\GTh_{\cC\cD}^{\bf 3875}
\;,
\non[1ex]
A_{2}^{I\dA}&=&
-\ft17\,\GG^J_{A\dA}\,\VV{\cC}{IJ}\VV{\cD}{A}\,\GTh_{\cC\cD}^{\bf 3875}
\;,
\non[1ex]
A_{3}^{\dA\dB}&=&
2\Gth\,\Gd_{\dA\dB}
+\ft1{48}\,\GG^{IJKL}_{\dA\dB}\,
\VV{\cC}{IJ}\VV{\cD}{KL}\,\GTh_{\cC\cD}^{\bf 3875}
\;.
\la{A123}
\ea
as can be shown by a rather lengthy computation which relies in
particular on the form of the projectors onto the irreducible parts of
its tensor products \cite{KoNiSa99a}. Thus, \Ref{crit} encodes the
complete set of consistency equations for the gauged theory, thereby
reducing the differential and bilinear identities for the a priori
unknown tensors $A_{1,2,3}$ to a single linear algebraic equation for
the embedding tensor $\GTh_{\cA\cB}$.  The formulation of this
equation as a projection condition allows to select the admissible
gauge subgroups of $E_{8(8)}$ by purely group theoretical
arguments. We shall not aim for a complete classification of the
solutions of \Ref{crit} here, but rather discuss the most important
examples.

The maximal compact subgroup of $E_{8(8)}$ satisfying the condition
\Ref{crit} is
\be\la{88}
SO(16)\supset G_0 = SO(8)\times SO(8) \;,
\ee
where the ratio of coupling constants of the two factors is
$g_1/g_2=-1$ such that the trace part $\Gth$ of $\GTh_{\cA\cB}$
vanishes.  The embedding tensor for $SO(8)\!\times\!SO(8)$ is
explicitly given by
\ba
&&\GTh_{IJ,KL} = \Gd\undersym{^{\vphantom{I}}_{I[K}
\,\varepsilon^{\vphantom{I}}_{L]J}\,}\;,
\quad 
\GTh_{IJ,A} =0=\GTh_{A,B}\;, \non
&&\mbox{with}\;\;
 \varepsilon^{\vphantom{I}}_{IJ}= 
{\rm diag} \,({\mathbb{I}}, -{\mathbb{I}})\;,
\la{T88}
\ea
with a $16\ra 8\!+\!8$ split of the indices $I, J$. This tensor
transforms in the ${\bf 135}$ under $SO(16)$ and automatically
satisfies \Ref{crit}, because there is no ${\bf 135}$ in the ${\bf
27000}$ of $E_{8(8)}$.
By contrast, no $SO(9)$ subgroup of $SO(16)$ satisfies \Ref{crit}, in
agreement with the fact that there is no $S^8$ compactification of
$D\!=\!11$ supergravity to three dimensions. Likewise, it may be shown
that apart from \Ref{88} none of its maximal compact subgroups of type
$SO(p)\!\times\!SO(16\!-\!p)$ satisfies \Ref{crit}.

Among the maximal subgroups of $E_{8(8)}$, we have found
three possible noncompact gauge groups
\ba
G_0 =E_{7(-5)}\times SU(2) 
&\quad\mbox{with}\quad& (g_{SU(2)}/g_{E_7})= -3 \;,\non
G_0 =E_{6(-14)}\times SU(3)  
&\mbox{with}& (g_{SU(3)}/g_{E_6})= -2 \;, \non
G_0 =F_{4(-20)}\times G_2  
&\mbox{with}& (g_{G_2}/g_{F_4})=-3/2 \;,
\la{nc}
\ea
all of which have nonvanishing trace $\Gth$. A further and
``extremal'' solution to \Ref{crit} is $G_0 =E_{8(8)}$ with
$\GTh_{\cA\cB}=\Gth\, \eta_{\cA\cB}$.  

An immediate question concerns the possible existence of AdS$_3$
ground states with additional symmetries. The fact that $d\!=\!11$
supergravity does not appear to admit compactifications of this type
indicates that the potential \Ref{potential} does not possess fully
supersymmetric stationary points \ci{Warner}. Inspection of the
supersymmetry variations \Ref{susyf} and \Ref{ferm} reveals that an
$N\!=\! 16$ invariant AdS$_3$ ground state can only exist if
$A_1^{IJ}\propto \delta_{IJ}$, i.e.\ if the trace of $\GTh_{\cA\cB}$
does not vanish.  Therefore the $SO(8)\!\times\!SO(8)$ gauged theory
\Ref{88} indeed cannot have a fully supersymmetric ground state;
rather, its maximally symmetric ground state is characterized by
$\cV={\mathbb{I}}_{248}$, which by \Ref{A123} implies
\be
A_{1}^{IJ}= -\ft17\,\varepsilon^{\vphantom{I}}_{IJ}\;,\quad
A_2^{I\dA} = 0 \;,\quad
A_3^{\dA\dB} = 0 \;, 
\ee
with $\varepsilon^{\vphantom{I}}_{IJ}$ from \Ref{T88}. The Lagrangian
\Ref{L} then reduces to the $(8,8)$ CS-theory of \cite{AchTow86} and
reproduces the vacuum found there. On the other hand, the truncation
of the gauge group $SO(8)\!\times\!SO(8)$ to its diagonal subgroup is
presumably related to the theory obtained by direct reduction of the
$D\!=\!4$ gauged theory, which no longer exhibits an $SO(8)$ invariant
AdS but rather a domain wall ground state \cite{LuPoTo97}.
Pursuing an analysis of the full theory along the lines of
\cite{FGPW99} should allow one to understand in detail the presumed
relation to the matrix string theory \cite {DeVeVe97} on the
boundary.

Finally, dimensional reduction of the gauged theories \Ref{L} yields
maximal gauged supergravities in two dimensions as well. In this
reduction, the vector fields ${B_\mu}^m$ give rise not only to
two-dimensional vectors, but also to the dual scalar fields at the
bottom of the infinite tower of dual potentials generated by the
$E_{9(9)}$ symmetry of $N\!=\!16 , D\!=\!2$ supergravity.  It would be
most interesting to see whether our construction can be extended to
yield gauged supergravities in two dimensions with
infinite-dimensional gauge groups.

A detailed account of our results will be presented elsewhere. This
work was supported in part by the European Union under Contract
No. HPRN-CT-2000-00122.


\begin{thebibliography}{10}

\bibitem{AGMOO00}
O. Aharony {\em et al.}, Phys. Rept. {\bf 323}, 183 (2000).

\bibitem{deWNic82}
B. de~Wit and H. Nicolai, Nucl. Phys. {\bf B208},  323  (1982).

\bibitem{SalSez89}
{\em Supergravities in diverse dimensions}, edited by A. Salam and E. Sezgin
  (World Scientific, Singapore, 1989).

\bibitem{Juli83}
B. Julia,  in {\em Unified field theories in more than 4 dimensions}, 
  edited by
  V.~D. Sabbata and E. Schmutzer (World Scientific, Singapore, 1983), pp.\
  215--236.

\bibitem{MarSch83}
N. Marcus and J. Schwarz, Nucl. Phys. {\bf B228},  145  (1983).

\bibitem{CrJuSc78}
E. Cremmer, B. Julia, and J. Scherk, Phys. Lett. {\bf 76B},  409  (1978).

\bibitem{DeVeVe97}
R. Dijkgraaf, E. Verlinde and H. Verlinde,
Nucl. Phys.  {\bf B500}, 43 (1997).

\bibitem{AchTow86}
A. Ach{\'u}carro and P.~K. Townsend, Phys. Lett. {\bf B180},  89  (1986).

\bibitem{LuPoTo97}
H. Lu, C.~N. Pope, and P.~K. Townsend, Phys. Lett. {\bf B391},  39  (1997).

\bibitem{CvLuPo00}
M. Cveti\v{c}, H. Lu, and C.~N. Pope, Phys. Rev. {\bf D62},  064028  (2000).

\bibitem{DKSS00}
N.S. Deger , A. Kaya, E. Sezgin and P. Sundell,
Nucl.. Phys. {\bf B573}, 275 (2000).

\bibitem{Hull84}
C.~M. Hull, Phys. Lett. {\bf B142},  39  (1984).

\bibitem{KoNiSa99a}
K. Koepsell, H. Nicolai, and H. Samtleben, JHEP {\bf 04},  023 (1999). 

\bibitem{KoNiSa99b}
K. Koepsell, H. Nicolai, and H. Samtleben, 
Class. Quant. Grav. {\bf 17},  3689 (2000).

\bibitem{CJLP98}
E. Cremmer, B. Julia, H. Lu, and C.~N. Pope, Nucl. Phys. {\bf B523},  73
  (1998).

\bibitem{deWNic84}
B. de~Wit and H. Nicolai, Nucl. Phys. {\bf B231},  506  (1984).

\bibitem{GuRoWa86}
M. G{\"u}naydin, L.~J. Romans, and N.~P. Warner, Nucl. Phys. {\bf B272},  598
  (1986).

\bibitem{Warner} 
This possibility was first suggested by N.P.~Warner.

\bibitem{FGPW99}
D.~Z.~Freedman, S.~S.~Gubser, K.~Pilch and N.~P.~Warner,
hep-th/9904017.
%%CITATION = HEP-TH 9904017;%%
%\end{thebibliography}

\end{thebibliography}
\end{document}